\documentclass[journal=jcisd8]{achemso}
\usepackage{array}
\usepackage{graphicx} 
\usepackage{xcolor}
\usepackage[utf8]{inputenc}
\usepackage{multirow}
\usepackage{amsmath}
\usepackage[hidelinks=true,colorlinks=true,linkcolor=teal, citecolor=blue,breaklinks=true]{hyperref}
\usepackage{xcolor}

\usepackage{longtable}
\usepackage{subcaption}  
\usepackage{float}
\usepackage{textcomp}
\usepackage{textopo}
\usepackage{booktabs}
\usepackage{amssymb}
\usepackage{csquotes}

\newcommand{\ptmpsi}{\textsc{PTM-P}si}

\title{\ptmpsi\ on the Cloud: A Cloud-compatible Workflow for Scalable, High-throughput
Simulation of Post-translational Modifications in Protein Complexes}

\author{Suman Samantray$^1$}
\affiliation{Physical Sciences Division, Pacific Northwest National Laboratory, Richland WA 99354}
\author{Margot Lockwood$^1$}
\affiliation{Environmental Molecular Sciences Laboratory, Pacific Northwest National Laboratory, Richland, WA 99352, USA}
\author{Amity Andersen}
\affiliation{Environmental Molecular Sciences Laboratory, Pacific Northwest National Laboratory, Richland, WA 99352, USA}
\author{Hoshin Kim}
\affiliation{Physical Sciences Division, Pacific Northwest National Laboratory, Richland WA 99354}
\author{Paul Rigor}
\affiliation{Center for Cloud Computing, Pacific Northwest National Laboratory, Richland, Washington 99354, USA}
\author{Margaret S. Cheung}
\affiliation{Environmental Molecular Sciences Laboratory, Pacific Northwest National Laboratory, Richland, WA 99352, USA}
\email{margaret.cheung@pnnl.gov}
\author{Daniel Mejia-Rodriguez}
\affiliation{Physical Sciences Division, Pacific Northwest National Laboratory, Richland WA 99354}
\email{daniel.mejia@pnnl.gov}

\date{September 2024}

\begin{document}

\maketitle
\begin{abstract}

We developed an advanced computational {framework} to accelerate {the study of the impact of} post-translational modifications on protein structures and interactions (\ptmpsi) using asynchronous, loosely coupled workflows on the Azure Quantum Elements Cloud platform. We seamlessly integrate emerging cloud computing assets that further expand the scope and capability of \ptmpsi\ {Python package} by refactoring it into a cloud { compatible} library. We employed a {\enquote{workflow of workflows} approach wherein a parent workflow spawns one or more child workflows, managing them, and acting on their results}. This approach enabled us to optimize resource allocation according to each workflow's needs,  and allowed us to use the cloud heterogeneous architecture for the computational investigation of a combinatorial explosion of thiol protein PTMs on an exemplary protein megacomplex critical to the Calvin-Benson cycle of light-dependent sugar production in cyanobacteria.  With \ptmpsi\ on the cloud, we transformed the pipeline for the thiol PTM analysis to achieve high throughput by leveraging the strengths of the cloud service.  \ptmpsi\ on the cloud reduces operational complexity and lowers entry barriers to data interpretation with structural modeling for a redox proteomics mass spectrometry specialist.

\end{abstract}
\section{Introduction}


Protein post-translational modifications (PTMs) are chemical modifications of amino-acid side chains that occur after biosynthesis\cite{Walsh2005}. Although proteins can be modified pre-, co-, and post-translationally, all these modifications are commonly referred to as PTMs, because most of them are made post-translationally, after the protein is folded.  A wide variety of possible combinations of oxidation-reduction (redox) PTMs on thiol-containing proteins shifts redox homeostasis to either oxidative or reductive stress, a change commonly associated with pathological states \cite{zhang_ajpcell2021}.
Thiol PTMs, involving covalently bonded thiols on cysteines, impact protein structures, molecular interactions, and stability\cite{paulsen_acschembiol2010}, depending on their positions and chemical types.  These thiol PTMs are dominantly reversible reactions in changing environments as an important mechanism for microbes to respond to light-dependent redox perturbations rapidly \cite{guo_mcp2014}. 

Since its inception, thiol-based redox proteomics with mass spectrometry, in which all protein thiols are subject to a variety of PTMs, has become the primary technology for broad identification and quantification of site-specific redox PTMs to understand the redox regulation in signaling and metabolism under physiological or pathological conditions\cite{zhang_ajpcell2021}. Profiling thiol redox PTMs and understanding their impact on biological functions is an extremely challenging task because of the combinatorial explosion of possible outcomes. 
Recent developments in deep learning have advanced the annotation and prediction of PTMs on proteins.\cite{shrestha2024post,ramazi2021post}
{However, the impact of PTMs on protein function and dynamics remains unclear without mechanistic insights from a structural model.} 
To meet the technical needs of data analysis and interpretation with structural models, we developed a user-friendly Python-based package, \textbf{P}ost-\textbf{t}ranslational \textbf{M}odifications on \textbf{P}rotein \textbf{S}tructure, and their \textbf{I}mpacts on dynamics and functions (\ptmpsi), that parses the complex workflow for numerous combinations of PTMs on proteins. \ptmpsi\ integrates several open-source software packages into its workflows, including software to infer protein structure from sequence (AlphaFold2\cite{jumper2020alphafold}), force field development for non-standard amino acids (including redox PTM forms) using quantum chemistry (NWChem\cite{apra2020nwchem}), free energy methods such as thermodynamic integration through molecular dynamics (GROMACS\cite{abraham2015gromacs}), and bound complex docking algorithms (AutoDock\cite{morris2008using}). 
In addition, the package uses a visual tool from ChimeraX\cite{pettersen_protsci2020}, aiming for an accurate interpretation of experimental data from thiol-based redox proteomics.\cite{mejia2023ptm}

{ In our previous work,} we used \ptmpsi\ to modify cysteines of the homotetramer of glyceraldehyde-3-phosphate dehydrogenase (NADP-GADPH, GAPDH, or GAP2) from \textit{Synechococcus elongatus} PCC 7942, as an example to interpret a sequence-structure-function relationship derived from thiol redox proteome data\cite{mejia2023ptm}. More specifically, we introduced the \textit{S}-nitrosothiol (SNO) group through a modification that can be named \enquote{\textit{S}-nitrosylation} or \enquote{\textit{S}-nitrosation}, depending on the specific reaction mechanism and species involved\cite{heinrich_brjpharm2013}. It was observed that, despite the minimal impact of SNO moieties on the global GAPDH conformation, the presence of an SNO group near the protein surface at cysteine 78 (CYS78) could allosterically inhibit the dehydrogenation reaction occurring at the internal catalytic site, specifically including cysteine 155 (CYS155). This inhibition mechanism is predicated on a reduction in the affinity for the cofactor NADP$^+$, attributed to steric repulsion between the SNO group and the adenosine 2'-phosphate, as well as the elimination of a hydrogen bond between the free thiol of CYS78 and the same adenosine 2'-phosphate.\cite{mejia2023ptm}.  
We were excited by the new insights from chemistry, modeling, and simulations brought by \ptmpsi\ in protein sciences, including oligomeric regulations.\cite{li2024deciphering}
%
We were eager to further our knowledge about the role that thiol PTMs play in the formation {and stability} of the protein megacomplex {composed of GAPDH, chloroplast protein 12 (CP12), and phosphoribulokinase (PRK)}. 
However, the constraints imposed by the computational resources of standard on-site clusters hindered our ability to scale up our protocol to obtain a predictive understanding of the regulatory mechanisms governed by redox PTMs { in an expedited timeframe}.

\begin{figure}[H]
    \centering
    \includegraphics[width=0.9\linewidth]{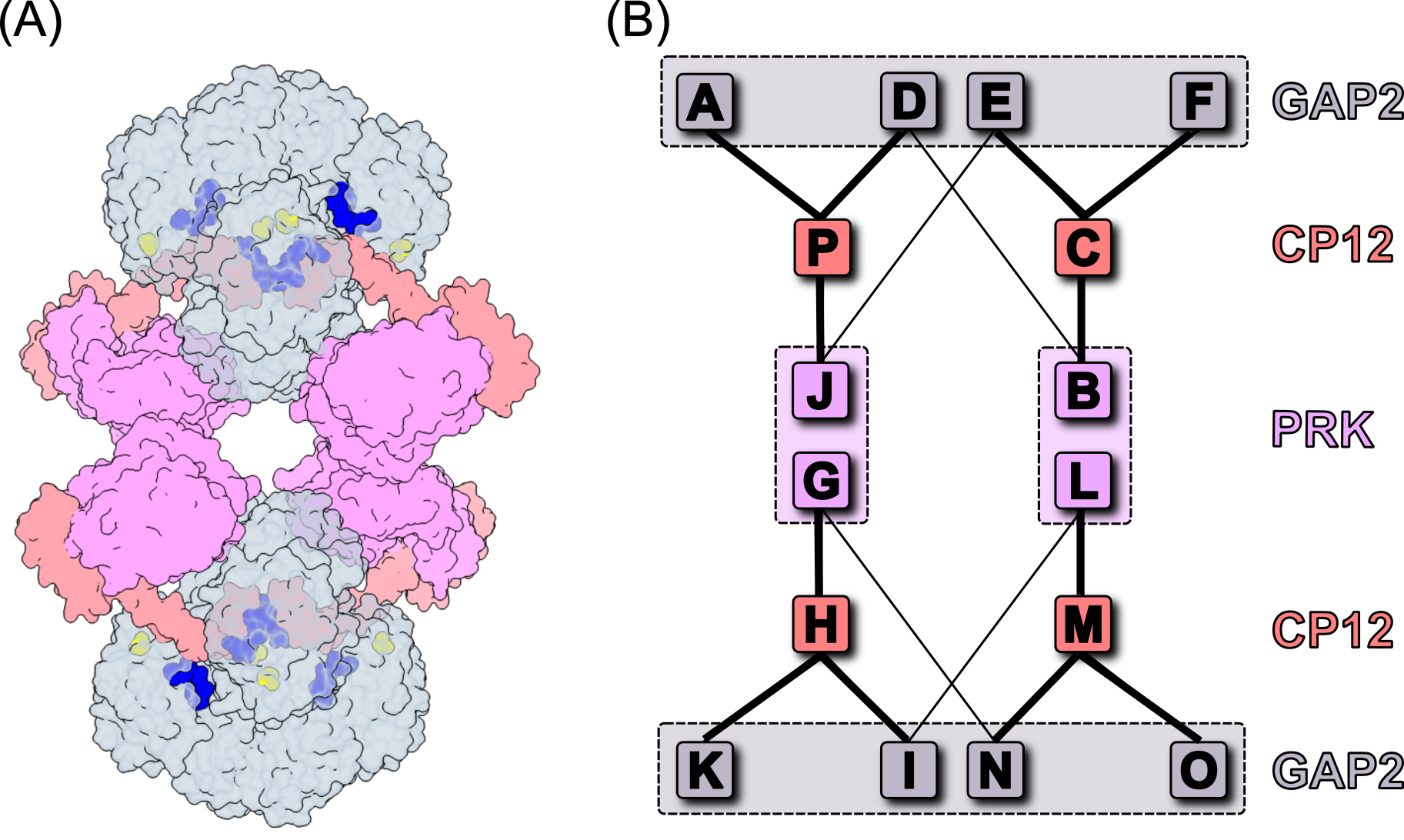}
    \caption{Structural representation of GAPDH/CP12/PRK megastructure, so-called the “dark complex” (PDB: 6GVE) (A), and a schematic representation of the interaction among the three modules (B). Gray represents GAPDH (or GAP2)  tetramers, orange represents CP12, magenta represents PRK dimers, blue represents the NAD cofactor, and yellow represents the cysteine targets for PTMs in this study. The capitalized letters in (B) are chain labels in the Protein Data Bank (PDB) file, and the line thickness represents the prevalence of the interaction in the simulations of 25 $\mu$s.}
    \label{fig:Dark_complex_schematic}
\end{figure}

The GAPDH/CP12/PRK \emph{dark complex} (Figure \ref{fig:Dark_complex_schematic}) is an important molecular redox switch of the Calvin-Benson cycle (CBC) in photosynthetic species--such as cyanobacteria, algae, and land plants--as the enzymatic activities of GAPDH and PRK are strongly inhibited when they form part of the ternary supramolecular GAPDH/CP12/PRK complex. In an oxidized state, the two C-terminal cysteines of CP12 form an intramolecular disulfide bridge that forces the intrinsically disordered protein into a hairpin conformation. The partially-folded C-terminal region of CP12 then binds to GAPDH in an obligatory intermediate step with a (GAPDH)-(CP12)$_2$ stoichiometry. An important prerequisite for the GAPDH/CP12 binary complex formation is the presence of NAD$^+$ (not NADP$^+$) inside the binding pocket of GAPDH. Two GAPDH/CP12 binary complexes then bind two oxidized PRK dimers via CP12's N-terminus, forming the dark complex spontaneously with a final (GAPDH)$_2$-(CP12)$_4$-(PRK)$_2$ stoichiometry \cite{marri2008spontaneous}. Although certain microorganisms possess a CP12 peptide sequence that allows the formation of an additional intramolecular disulfide bond near the N-terminus, the presence of this second disulfide bridge does not appear to be essential for the \emph{dark complex} formation and stability.
In a highly reduced state, the small scaffold protein CP12 becomes intrinsically disordered as a monomeric unit, thus impeding the formation of the supramolecular dark complex. Furthermore, the GAPDH/CP12/PRK complex dissociates in highly reducing environments in a sequence of steps that involve the reduction of the PRK disulfide bonds, the subsequent dissociation of PRK from the ternary complex, the reduction of the CP12 C-terminus disulfide bonds, and the final dissociation of the GAPDH/CP12 binary complex. In their free form, GAPDH and PRK resume enzymatic activity. The regulation mechanism exerted by reversible thiol PTMs on the formation and dissociation of the dark complex remains unclear. This regulation tunes the carbon flow between the photosynthetic reduction cycle in CBC and the oxidative pentose phosphate cycle, depending on the light intensity during the natural diel cycle  \cite{tamoi2005calvin}.

\begin{figure}[H]
    \centering
    \includegraphics[width=0.9\linewidth]{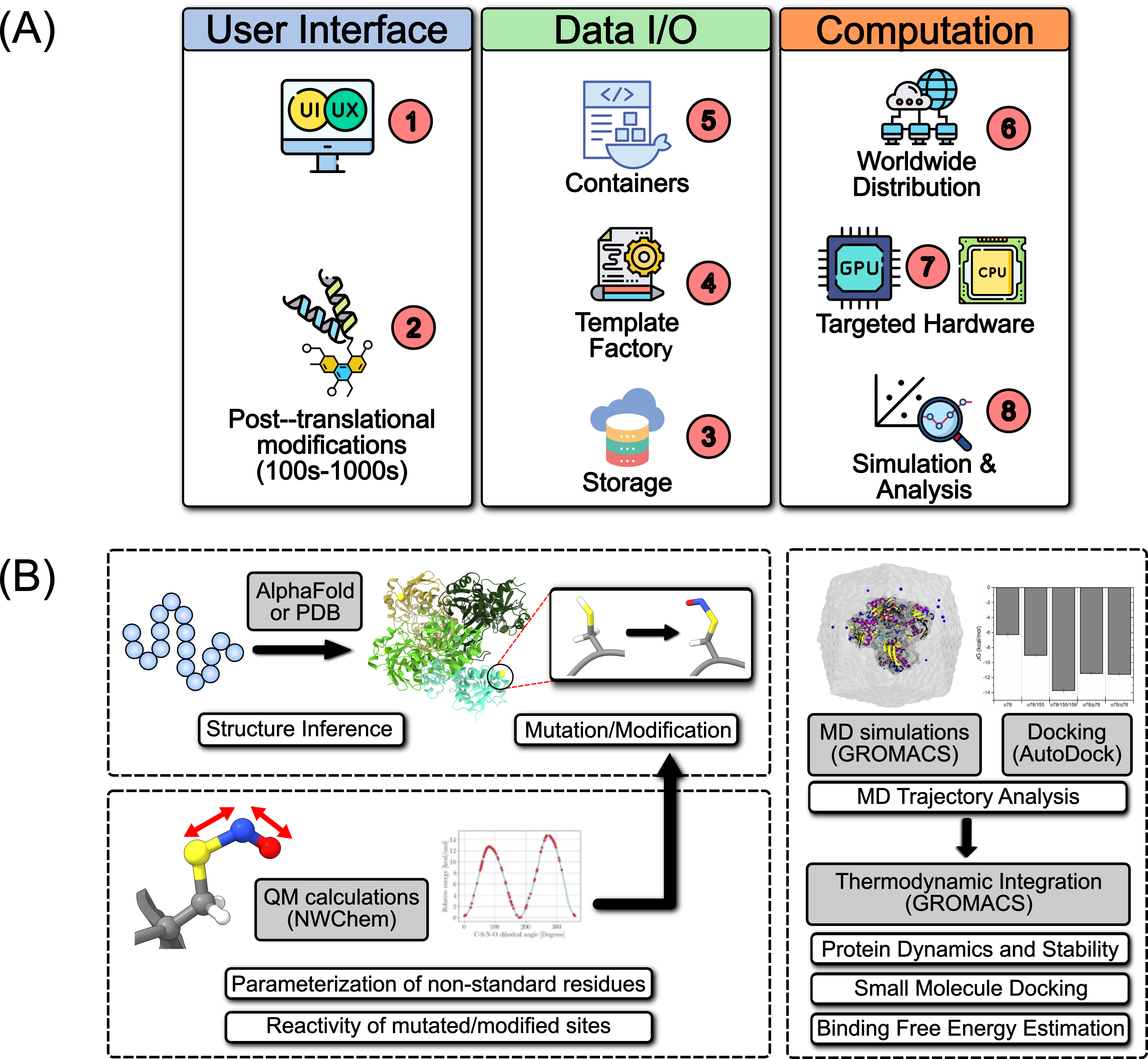}
    \caption{(A)Overview of \enquote{a flow of workflow} for standing up \ptmpsi\ on the cloud. 1) The user can interact through the Azure Quantum Elements Open OnDemand\texttrademark{} interface or the command line. 2) Once a protein target is identified, \ptmpsi\ either infers a three-dimensional structure by using AlphaFold2 or reads a PDB file. Then, the user can use \ptmpsi\ to generate hundreds to thousands of combinatorially PTMed structures.     
    3) These structures are then stored in a hierarchical file with unique identifiers. 4) The template factory then generates all scripts needed to follow the selected workflow. It interfaces with the appropriate hardware and software, including a queueing system, either SLURM or PBS. 5) All software called during the workflows is containerized using Apptainer images and automatically pulled from the container registry. 6) The user submits a top-level script to start the workflow, potentially distributed over several nodes worldwide. 7) The workflow is set up to take advantage of the best hardware available for the given task, considering the constraints of the software stack. 8) Depending on the selected workflow, \ptmpsi\  generates appropriate scripts that perform analyses on the trajectories from the production runs. It includes thermodynamic integration and PCA. (B) The workflow of \ptmpsi\ with the addition of a module for trajectory analysis.  }
    \label{fig:Overview_PTMPsi}
\end{figure}

To understand the fundamental mechanisms of the enzymatic activities in the CBC turnover, the redox proteome of \textit{S. elongatus} was measured by high-throughput redox mass spectrometry after light disturbance\cite{johnson_arxiv2025}. 
The first version of \ptmpsi\ was developed to provide mass spectrometry specialists with a structural interpretation of their datasets from computational models. However, this {version} could not efficiently {orchestrate} thousands of simulations stemming from a combinatorial way to modify all thiol groups that might participate in the day/night redox homeostasis of cyanobacteria\cite{ansong_frontiers2014}.
In response to the immense need for high-performance computing (HPC), we deployed \ptmpsi\ on the cloud with the approach of {\enquote{workflow of workflows}} on { a private preview version of} Microsoft's Azure Quantum Elements (AQE) cloud computing infrastructure, which we visually describe in Figure~\ref{fig:Overview_PTMPsi}A. To manage accompanying data output from these simulations, we also augmented the {package} with a tool for post-processing analyses of massive datasets { that comprises methods} such as principal component analysis (PCA), dimensionality reduction, clustering, and kinetic transition network (KTN). We also introduced a local vs. global scoring function to evaluate the PTM influence on structures and dynamics at a quick glance (Figure~\ref{fig:Overview_PTMPsi}B). 

We demonstrated the performance of \ptmpsi\ and its new functionality in a computational investigation of several strategically PTMed GAPDH/CP12/PRK dark complex systems from an experimentally resolved mega structure (PDB: 6GVE) of the related cyanobacteria \textit{Thermosynechococcus vestitus BP-1} {(NCBI:txid197221)}. By leveraging the unique advantages of cloud computing, we were able to obtain all simulation instances at a significantly faster rate than other HPC platforms that suffer from heavy usage and prolonged turnaround times.
The cysteines (equivalent to CYS78 in the previous study) of GAPDH  act like a linchpin that holds the megacomplex together. Their oxidation with sizable thiol groups pivots to destabilize the assembly of the dark complex, even without a significant change in the structure of individual subunits. Our  work demonstrates a scalable research capability of \ptmpsi\ 
on the cloud that enables a mechanistic understanding of protein sequence, structure, and dynamics related to redox homeostasis at a proteome level. 

\section{Methods}

\subsection{Cloud-based HPC cluster}

The simulations described here were carried out on the AQE platform, housed and developed by Microsoft Corporation, USA. 
AQE offers a hybrid computing environment with a wide variety of hardware for HPC { through Azure Virtual Machines (VMs)}. An AQE cluster { can be configured with several types of VMs, each optimized for a specific purpose.} Furthermore, each cluster includes utility nodes for scripting, analysis of results, rendering, and visualization through the Open OnDemand\texttrademark{} web-based client\cite{OpenOndemand}. Depending on the workflow characteristics, the compute nodes can be interconnected with fast fabrics such as InfiniBand or NVLink. For this research, we used two types of nodes: 1) graphics processing unit (GPU)-nodes containing 96 Intel Xeon (Sapphire Rapids) processor cores and eight H100 GPUs, connected with high-speed NVLink GPU interconnects, per node {(ND\_H100\_v5 series)}; and 2) CPU-only nodes containing 120 AMD EPYC 7V12 processor cores per node, {4 GB of RAM per core, and a 200 Gb/s Mellanox HDR InfiBand interconnect (HBv2 series). \ptmpsi\ was also tested on other Azure VM families, including the NCv3 series featuring V100 GPUs and the NC\_A100\_v4 series with A100 GPUs.}
The heterogeneous nature of the Cloud hardware architecture is suitable for diverse workloads and workflows. Furthermore, jobs from other traditional HPC platforms are easy to migrate to the AQE through a SLURM queuing system \cite{SLURM1,SLURM2}.

\subsection{Sample preparation}

We computationally investigated the impact of selected thiol PTMs on the GAPDH/CP12/PRK complex (\enquote{dark complex}) with \ptmpsi. We used the experimentally determined protein megacomplex from \textit{Thermosynechococcus vestitus BP-1}\cite{macfarlane_pnas2019} (PDBID: 6GVE).
We set the native complex from 6GVE as {Simulation (Sim.)}0 in Table \ref{table:PTM}. The NAD cofactor is included in each monomeric GAPDH, as found in the 6GVE structure. {Sim.}0 contains two disulfide bridges in PRK (CYX13-CYX35 and CYX224-CYS230) and two disulfide bridges in CP12 (CYX61-CYX70 and CYS19-CYS29). {We note that CYS77} (equivalent to CYS78 in our prior study) in GAPDH is a free cysteine.
We systematically oxidized cysteine 77 of GAPDH with \textit{S}-nitrosylation (SNO), \textit{S}-sulfhydration (SSH), and \textit{S}-glutathionylation (SSG) in {Sim.}1 to 3. 
We broke the two pairs of disulfide bridges in PRK in {Sim.}4 to 9 by introducing SNO, SSH, and SSG modifications in both cysteines involved in the original disulfide bridge. We did the same for CP12 in {Sim.} 10 to 15 in Table \ref{table:PTM}. 

For each { simulation} instance, the system is solvated with explicit water molecules in a truncated dodecahedron box with a 1 nm padding for all-atom molecular dynamics (MD) simulations, leading to approximately 800k atoms per instance. Enough Na$^+$ and Cl$^-$ ions are added to the simulation box to neutralize the net charge and achieve a 0.154 M salt concentration. Minimization then proceeds in a stepwise fashion by restraining the positions of the atoms in the protein complex with decreasing force constants. System equilibration also proceeds in a stepwise fashion. First, the positions of the protein complex are restrained while the solvent is equilibrated at 300 K for 500 ps in the canonical (NVT) ensemble using the stochastic velocity rescaling (V-rescale) thermostat \cite{bussi2007}, followed by equilibration at 300 K and 1 atm for 500 ps in the isothermal-isobaric (NPT) ensemble with the stochastic cell rescaling (C-rescale) barostat\cite{bernetti2020} in addition to the V-rescale thermostat. Then, the position restraints on the protein complex are lifted, and the NVT and NPT equilibrations are repeated. This ensures a smooth equilibration of the perturbed complex containing the redox PTMs.
\ptmpsi\ was set up to run production NPT molecular dynamics simulations at 300 K and 1 atm for 100 ns using the V-rescale thermostat and C-rescale barostat. All MD simulations constrained the bonds to hydrogen atoms using the LINCS algorithm\cite{hess1997} and used a time step of 2 fs.

\begin{table}[ht]
\centering
\begin{tabular}{r p{0.7\textwidth} l }
\toprule
\multicolumn{1}{l}{\textbf{{Sim.}\texttt{\#}}} & \textbf{\hspace{5em}Site (Residue)} & \textbf{PTM type} \\
\toprule
0 & \hspace{5em}PDB ID: 6GVE & as is\\[2ex]
\multicolumn{3}{c}{\textbf{GAP2: A, D, E, F, I, K, N, O }} \\
\midrule
1 & \hspace{5em}CYS77 & SNO \\
2 & \hspace{5em}CYS77 & SSH \\
3 & \hspace{5em}CYS77 & SSG \\[2ex]
\multicolumn{3}{c}{\textbf{PRK: B, L, G, J}} \\
\midrule
4 & \hspace{5em}CYX13, CYX35 & SNO \\
5 & \hspace{5em}CYX13, CYX35 & SSH \\
6 & \hspace{5em}CYX13, CYX35 & SSG \\
7 & \hspace{5em}CYX224, CYX230 & SNO \\
8 & \hspace{5em}CYX224, CYX230 & SSH \\
9 & \hspace{5em}CYX224, CYX230 & SSG \\[2ex]
\multicolumn{3}{c}{\textbf{CP12: P, C, H, M}} \\
\midrule
10 & \hspace{5em}CYX61, CYX70 & SNO \\
11 & \hspace{5em}CYX61, CYX70 & SSH \\
12 & \hspace{5em}CYX61, CYX70 & SSG \\
13 & \hspace{5em}CYX19, CYX29 & SNO \\ 
14 & \hspace{5em}CYX19, CYX29 & SSH \\ 
15 & \hspace{5em}CYX19, CYX29 & SSG \\ 
\bottomrule
\end{tabular}
\caption{
Each {simulation (Sim.)} number corresponds to a specific case with a particular set of PTMs in specific chain locations. \enquote{CYS} denotes a free cysteine residue, while \enquote{CYX} represents part of a disulfide bridge in PDB ID 6GVE (i.e. {Sim.}0{, and is the considered as the reference or baseline in this work}) -- there are two disulfide bridges in PRK and in CP12, respectively. Abbreviations of PTMs on a specific cysteine  include SNO (\textit{S}-nitrosylation), SSH (\textit{S}-sulfhydration), SSG or (\textit{S}-glutathionylation).}
\label{table:PTM}
\end{table}


\subsection{Workflow of Trajectory Analysis} \label{sec:analysis}

We added common molecular dynamics trajectory analysis tools to \ptmpsi\ and streamlined their input/output interface to ease data processing. The streamlined procedure saves time, reduces manual errors, and standardizes the analysis processes. Among these tools are simple post-processing steps (removal of water molecules or ions, elimination of rotational and translational degrees of freedom) provided by the GROMACS package. We generate configurations, followed by  
(i) a suite of univariate metrics analyses,  
(ii) reduction of data representations,  
(iii) multivariate metrics analysis, and  
(iv) kinetic transition network construction,  
as described in the subsections below.

\subsubsection{Univariate Metrics}

\paragraph{Radius of Gyration (\(R_g\))}  
Radius of gyration measures the compactness of the molecule. It is defined as  
\begin{equation}\label{eq:rg}
R_g(t) \;=\; \sqrt{\frac{1}{N_{\rm atom}}\sum_{k=1}^{N_{\rm atom}}\bigl\|\mathbf{r}_k(t)-\mathbf{r}_{\rm cm}(t)\bigr\|^2}\,,
\end{equation}  
where \(N_{\rm atom}\) is the total number of atoms in the complex, \(\mathbf{r}_k(t)\) is the position of atom \(k\) at time \(t\), and \(\mathbf{r}_{\rm cm}(t)\) is the center-of-mass position of all atoms.

\paragraph{Root Mean Square Deviation (RMSD)}  
RMSD quantifies the deviation of instantaneous atomic positions from a reference (average) conformation:  
\begin{equation}\label{eq:rmsd}
\mathrm{RMSD}(t)
= \sqrt{\frac{1}{N_{\rm atom}}\sum_{k=1}^{N_{\rm atom}}\bigl\|\mathbf{r}_k(t)-\mathbf{r}_k^{\rm ref}\bigr\|^2}\,,
\end{equation}  
where \(\mathbf{r}_k^{\rm ref}\) is the position of atom \(k\) { in the reference state}.

\paragraph{Root Mean Square Fluctuation (RMSF)}  
RMSF describes per-residue flexibility. For residue \(i\), using { either} backbone heavy-atom center-of-mass { or C$_\alpha$ position as} \(\mathbf{R}_i(t)\),  
\begin{equation}\label{eq:rmsf}
\mathrm{RMSF}_i
= \sqrt{\frac{1}{T_{\rm frame}}\sum_{t=1}^{T_{\rm frame}}\bigl\|\mathbf{R}_i(t)-\langle\mathbf{R}_i\rangle\bigr\|^2}\,,
\end{equation}  
where \(T_{\rm frame}\) is the total number of frames and \(\langle\mathbf{R}_i\rangle\) is the time average of \(\mathbf{R}_i(t)\).

\paragraph{Secondary Structure}  
Secondary structure assignment tracks helix, sheet, and coil fractions over time using the Dictionary of Secondary Structure in Proteins (DSSP) algorithm\cite{kabsch1983,Joosten2010}.

\subsubsection{Reduction of Data Representations} 

A user may select representative groups of residues to balance computational cost and interpretability. We focus on three regions:  
(i) {PTM shell}: all atoms of the modified residue plus neighbors within 3.0~\AA;
(ii) {NAD shell}: all NAD/NADH atoms plus neighbors within 3.0~\AA;(iii) {Interface atoms}: any atom within 2.0~{\AA} of CP12 at the interface of GAP2 and PRK.

\subsubsection{Multivariate Analysis}

Multivariate analysis refers to the statistical analysis of data that involves multiple variables or \enquote{features}. This type of analysis is essential when dealing with complex and massive datasets where each observation consists of multiple measurements. We deployed multivariate analysis to understand the relationships between the variables in the datasets. We used the technique of principal component analysis (PCA) to reduce the dimensionality of the datasets by transforming them into a new set of orthogonal variables that capture the most variance.
By pooling measurements per frame into one feature matrix, we  
revealed correlated motions invisible to one-dimensional plots by applying PCA, { reducing} the redundancy among overlapping metrics, and transforming the {data} from all instances under various PTM perturbations into a common coordinate system\citep{samantray2021influences}. 
{We ranked the impact of PTM perturbations on the conformational ensemble in the simulations by ranking the variability obtained through PCA.}

\textbf{Feature extraction: }For each simulation system \(i\) in Table \ref{table:PTM}, there are \(N\approx1000\) samples. At timeframe \(t\), we compute an \(M\)-dimensional feature vector from a selection of atoms that is either from the whole protein complex or from the reduced representation as described in the previous subsection:
\begin{align}
x^{t}_{i}
= \bigl[R_g,\;\mathrm{RMSD},\;B,\;\phi,\;\psi,\;
\mathrm{SS}_{\mathrm{helix}},\;
\mathrm{SS}_{\mathrm{sheet}},\;
\mathrm{SS}_{\mathrm{coil}},\;
\mathrm{n}_{\mathrm{PTM\text{-}Residue}},\;\dots\bigr]^t_i
\in\mathbb{R}^M.
\end{align}
Here, 
\(R_g\) and \(\mathrm{RMSD}\) are measured in nm. \(B\) is the average B-factor measured in nm\(^2\), calculated by
  \begin{equation}\label{eq:Bfactor}
  B = \frac{8\pi^2}{3}\,\bigl\langle\mathrm{RMSF}^2\bigr\rangle,
  \end{equation}
  where \(\langle\mathrm{RMSF}^2\rangle\) denotes the average squared RMSF per residue in a selected representation.
 The \(\phi\) and \(\psi\) are the average backbone dihedral angles expressed in radians.
 $\mathrm{SS}_{\mathrm{helix}}$, $\mathrm{SS}_{\mathrm{sheet}}$, 
 and 
 $\mathrm{SS}_{\mathrm{coil}}$ are the fractional secondary-structure propensity.
 The \(\mathrm{n}_{\mathrm{PTM\text{-}Residue}}\) denotes the count of residual contacts between the PTMed residue and its neighboring residues within 3.0~\AA.

Stacking these vectors as transposes:
\begin{equation}\label{eq:Xi}
X_i^\mathsf{T}
=\begin{pmatrix}
x_{i,1}^\mathsf{T}\\
x_{i,2}^\mathsf{T}\\
\vdots\\
x_{i,N}^\mathsf{T}
\end{pmatrix}
\in\mathbb{R}^{N\times M},
\end{equation}  
and concatenating all of 16 simulation {instances} (e.g., {Sim.}0 to 15 in Table \ref{table:PTM}):
\begin{equation}\label{eq:X}
X
=\begin{pmatrix}
X_1\\
X_2\\
\vdots\\
X_{16}
\end{pmatrix}
\in\mathbb{R}^{16N\times M}.
\end{equation}  
We mean-center and scale each column, then form the pooled covariance matrix,  
\begin{equation}\label{eq:C}
C = \frac{1}{16N-1}\,X^\mathsf{T}X.
\end{equation}  
PCA solves the eigenproblem below:  $v_j$ is an eigenvector and $\lambda_j$ is an eigenvalue of a principal component $j$. Principal Component 1 (PC1) has the largest eigenvalue of the covariance matrix. The eigenvalues are defined as 
\begin{equation}\label{eq:eig}
C\,v_j = \lambda_j\,v_j,\quad \lambda_1\ge\lambda_2\ge\cdots\ge0,
\end{equation}  
and we retain the {first} \(K\) number of principal components such that the cumulative variance preserves, {at least, 90\% of the total variance},
\begin{equation}\label{eq:var}
\frac{\sum_{j=1}^K\lambda_j}{\sum_{j=1}^M\lambda_j}\ge0.9.
\end{equation}

\textbf{Ranking PTM perturbations:} the mean projection onto the first principal component (PC1) for {Sim.}\(i\) is given by
\begin{equation}\label{eq:PC1}
\langle\mathrm{PC1}\rangle_i
= \frac{1}{N}\sum_{t=1}^N(X_i\,v_1)_t
\end{equation} 
and \begin{equation}\label{eq:delPC1}
\quad
\Delta\langle\mathrm{PC1}\rangle_i
= \langle\mathrm{PC1}\rangle_i-\langle\mathrm{PC1}\rangle_{\rm 0}.
\end{equation} 
 measures the PTM perturbations on the conformational ensemble from {Sim.}\(i\)
 along the dominant mode of variation relative to those of {Sim.}0. The magnitude \(|\Delta\langle\mathrm{PC1}\rangle_i|\)  indicates deviation from the unmodified reference ensemble of {Sim.}0, and the sign reflects the direction of the shift in collective conformational space. Simulation systems are ranked by $\Delta\langle\mathrm{PC1}\rangle_i$.  

The \emph{sign} of an eigenvector component from $\langle\mathrm{PC1}\rangle_i$ informs the importance of individual attributed features: a positive score arises from positive loadings $\ell_k$ in $v_1$; likewise, negative. 

\begin{equation}\label{eq:PC1score}
\langle\mathrm{PC1}\rangle_i
= \sum_{k=1}^M \ell_k\,\bar x_{i,k}
= \mathbf{v}_1^{\mathsf T}\,\bar{\mathbf x}_i
\end{equation}
Here, $v_1 = [\ell_1,\ell_2,\dots,\ell_M]^{\mathsf T}$ and each loading $\ell_k$ informs the strength of the feature $k$ driving first principal component PC1 in either a positive “+” or a negative “–” direction. 
 
\subsubsection{Kinetic Transition Network}

The kinetic transition network (KTN) formalism transforms a high-dimensional, noisy molecular trajectory into a simplified yet informative network model that captures both the \textit{thermodynamics} (through stationary distributions) and \textit{kinetics} (through transition rates and mean first passage time (MFPTs)) of the system. 
We employed KTN -- a graph-based model\citep{samantray2021effects}-- to
estimate transition probabilities between metastable states in a conformational ensemble and to identify representative states for each basin.
It captures the dynamic connectivity for visualizing and quantifying the kinetics of conformational switching between basins.
For this study, we reduced the dimensionality of its feature matrix \(X \in \mathbb{R}^{N \times M}\) from Equation
\eqref{eq:Xi}
by projecting each frame onto the top two principal components. This step transforms the high-dimensional feature space into an interpretable two-dimensional subspace that preserves the most dominant collective motions:
\begin{equation}\label{eq:S_expanded}
S = X[v_1\;\,v_2] \in \mathbb{R}^{N\times 2},
\end{equation}
where \(v_1\) and \(v_2\) are the first and second eigenvectors of the global covariance matrix \(C\), and \(S_t = (s_{t,1}, s_{t,2})\) denotes the two-dimensional projection of the time frame \(t\). This projection provides a kinetically meaningful embedding where similar conformations lie close to each other.

Next, we use \(k\)-means clustering to partition this projected trajectory into \(k\) distinct kinetic microstates, each representing a metastable basin in the conformational landscape. The objective is to minimize the within-cluster variance, grouping similar projections together:
\begin{equation}\label{eq:kmeans_expanded}
\{\mu_j\} = \arg\min_{\{\mu\}} \sum_{t=1}^N \min_{0 \le j < k} \|S_t - \mu_j\|^2,
\end{equation}
where \(\mu_j \in \mathbb{R}^2\) is the centroid of cluster \(j\), and \(k\) is a user-defined parameter for defining the number of clusters (typically \(k=5\)) representing the number of coarse-grained states. Each frame \(t\) is assigned a discrete cluster label \( m_t \in \{0, 1, \dots, k-1\}\), identifying the kinetic basin to which it belongs.

To quantify the kinetics of transitions between these states, we construct a \textit{Markov transition matrix} \(T \in \mathbb{R}^{k \times k}\), which encodes the conditional probability of moving from state \(i\) to state \(j\) in a single timestep:
\begin{equation}\label{eq:T_expanded}
T_{ij} = \frac{\#\{t \mid \ell_t = i,\; \ell_{t+1} = j\}}{\#\{t \mid \ell_t = i\}}.
\end{equation}
Here, the numerator counts how often the system transitions from cluster \(i\) to cluster \(j\) in consecutive frames, and the denominator normalizes by the total number of times the system is observed in cluster \(i\). This yields a \textit{row-stochastic matrix}, where each row sums to 1.

The stationary distribution \(\displaystyle \pi = [\pi_0,\dots,\pi_{k-1}]\) is obtained as the normalized left eigenvector of \(T\),
\[
  \pi = \pi\,T,
  \quad
  \sum_{j=0}^{k-1}\pi_j = 1.
\]
In our KTN figures, the node area scales with \(\pi_j\), and the label “\(p\)” on each node is the stationary probability \(\pi_j\).

We then compute the \textit{stationary distribution} \(\pi = [\pi_0, \dots, \pi_{k-1}]\), satisfying the fixed-point condition \(\pi = \pi T\), which reflects the probability of finding the system in each cluster at a steady state. To compute kinetic observables beyond immediate transitions, we form the \textit{fundamental matrix}:
\begin{equation}\label{eq:Z_expanded}
Z = \left(I - T + \mathbf{1} \pi^\mathsf{T} \right)^{-1},
\end{equation}
where \(I\) is the identity matrix and \(\mathbf{1}\) is a column vector of ones. This matrix quantifies cumulative occupancy and paths in the network and forms the basis for computing \textit{mean first-passage time (MFPT)} and \textit{mean residence time}, $\tau$.

The MFPT from state \(i\) to state \(j\), denoted \(\mathrm{MFPT}_{i \to j}\), is the expected number of steps required to reach \(j\) starting from \(i\), and is computed as:
\begin{equation}\label{eq:mfpt_expanded}
\mathrm{MFPT}_{i \to j} = \frac{Z_{jj} - Z_{ij}}{\pi_j}.
\end{equation}
This quantity measures \textit{kinetic separation} between basins; that is, how difficult it is to transition from one state to another under the Markovian dynamics.

The \textit{mean residence time} \(\tau_j\) for each state \(j\) quantifies how long, on average, the system stays within a state before transitioning out:
\begin{equation}\label{eq:tau_expanded}
\tau_j = \frac{1}{1 - T_{jj}}.
\end{equation}
This value is particularly important for identifying deep kinetic traps, which correspond to basins with large self-transition probabilities \(T_{jj}\). All temporal quantities (e.g., MFPT, \(\tau_j\)) are scaled by the  time step (i.e., \(\Delta t = 0.1\)~ns) to convert frame units into real time in a physical world.

For each cluster \(j\), we identify a \textit{representative structure} \(t_j^*\) that best exemplifies the conformational features of that basin. This is selected where the minimum of the energy landscape is closest to the cluster centroid:
\begin{equation}\label{eq:rep_expanded}
t_j^* = \arg\min_t \|S_t - \mu_j\|.
\end{equation}
This representative structure is used for the visualization and interpretation of the metastable states identified by the KTN.

\subsection{Thermodynamic Integrations}

While PCA and KTN characterize how PTMs alter the conformational ensemble and kinetics of the protein, they do not directly quantify the energetic cost of introducing a given PTM. To complement these structural and dynamical insights, we perform \textbf{thermodynamic integration (TI)}—a rigorous method from statistical mechanics used to compute the free-energy difference \(\Delta G\) between the unmodified reference system and its PTM-perturbed counterpart.

Thermodynamic integration introduces a continuous coupling parameter \(\lambda \in [0,1]\) that interpolates between two Hamiltonians, \(U(0)\) corresponding to the unmodified reference system and \(U(1)\) representing the fully perturbed (PTMed) system. For a fixed \(\lambda_k\), the system is equilibrated and sampled under the hybrid potential \(U(\lambda_k)\), which smoothly blends contributions from both endpoints. The free-energy derivative with respect to \(\lambda\), given by the ensemble average of the Hamiltonian gradient,
\begin{equation}\label{eq:dUdlam}
\left\langle \frac{\partial U}{\partial \lambda} \right\rangle_{\lambda_k},
\end{equation}
is evaluated at discrete \(\lambda_k\) points across the interval [0,1], typically spaced uniformly.

The total free-energy difference \(\Delta G\) is then obtained by numerically integrating the averaged gradients across the entire alchemical path:
\begin{equation}\label{eq:TI}
\Delta G = \int_0^1 \left\langle \frac{\partial U}{\partial \lambda} \right\rangle_\lambda \, d\lambda.
\end{equation}
In practice, we perform short simulations (typically 2~ns) at each \(\lambda_k\), extract the instantaneous \(\partial_\lambda U\), and average them to obtain the required integrand. The integration is carried out using methods such as trapezoidal or Simpson’s rule depending on the resolution and symmetry of the \(\lambda\)-schedule.

To improve statistical efficiency and reduce integration bias, particularly when end-point fluctuations are asymmetric, we apply \textbf{Bennett’s Acceptance Ratio (BAR)} method. This estimator exploits the overlap in configuration space between adjacent \(\lambda\) ensembles to more accurately reconstruct \(\Delta G\) from bidirectional samples:
\begin{equation}\label{eq:BAR}
\Delta G = -k_B T \ln
\frac{
\left\langle f\left[U(1) - U(0) + C\right] \right\rangle_{0}
}{
\left\langle f\left[U(0) - U(1) - C\right] \right\rangle_{1}
},
\quad
f(x) = \frac{1}{1 + e^{\beta x}},
\end{equation}
where \(f(x)\) is the Fermi function, \(\beta = 1/k_B T\) is the inverse thermal energy, and \(C\) is a self-consistently determined shift to minimize the variance of the free-energy estimate. The averages \(\langle \cdot \rangle_0\) and \(\langle \cdot \rangle_1\) denote sampling over the unmodified and modified ensembles, respectively.

Together, TI and BAR provide a rigorous thermodynamic complement to our structural and kinetic analyses, allowing us to assign energetic penalties or stabilizations associated with PTM-induced alterations. These energetic values serve as a quantitative foundation for interpreting how different PTMs influence protein function, binding, or stability.
{To calculate relative binding free energy (RBFE), we considered a closed thermodynamic cycle composed of two alchemical transformations:}
\begin{equation}\label{eq:RBFE}
{\Delta\Delta G_{RBFE} (A\xrightarrow{}B) = \Delta G_{protein} (A\xrightarrow{}B) - \Delta G_{water} (A\xrightarrow{}B)}
\end{equation}
{where \(\Delta G_{protein} (A\xrightarrow{}B)\) and \(\Delta G_{water} (A\xrightarrow{}B)\) represent the transformation from the PTMed cysteine (A) to its regular form (B) in protein complex and bulk water, respectively. For the transformation in bulk water, \(\Delta G_{water} (A\xrightarrow{}B)\), we simulated both the cysteine and its glutathionylated form, capped with an acetyl group (N-terminus) and N-methyl group (C-terminus) in TIP3P water. The same simulation protocols used for the protein systems were applied and TI calculations were performed to obtain \(\Delta G_{water} (A\xrightarrow{}B)\). For each transformation, we employed the dual topology approach, in which the parameters of the PTMed cysteine in state A were gradually transformed into those of a regular cysteine in state B. Specifically, the sulfur atom from glutathione, covalently bound to the cysteine sulfur, was converted into a hydrogen atom bound to the cysteine sulfur during the alchemical transformation, while the remaining atoms of glutathione were converted into dummy atoms that retained the mass of a hydrogen atom but had no non-bonded interactions. All bonded and non-bonded parameters for state A (PTMed cysteine) were derived using NWChem within the \ptmpsi\ package.\cite{mejia2023ptm} This approach has been applied to investigate free energy differences in similar contexts, including point mutations of amino acids in protein complexes,\cite{katzTI-mut1,JespersTI-mut2,AldeghiTI-mut3} pKa estimations of protonated amino acids,\cite{WilsonTI-pka} and ligand binding energies in protein binding pockets.\cite{AthanasiouTI-lig,KimTI-lig} Comprehensive details of the TI protocol and parameterization procedures have been reported in our previous study.\cite{mejia2023ptm}}


\section{Results and Discussion}

{\subsection{Softwarization: Making \ptmpsi\ cloud compatible} }
\ptmpsi\ is a Python package used to facilitate the computational investigation of PTMs on protein structures and their impacts on dynamics and functions in a workflow that uses well-established software, including AlphaFold2, NWChem, AutoDock, and GROMACS (Figure~\ref{fig:Overview_PTMPsi} B). { The first version of \ptmpsi\ assumed that the user was targeting a homogeneous HPC cluster with no GPU accelerators available. To make \ptmpsi\ a cloud-compatible package, we implemented the \enquote{workflow of workflows} approach, wherein a parent workflow triggers one or more child workflows, manages them, and acts on their results.} The refactoring into a cloud { compatible} package { results in} optimized resource allocation tailored to the computational needs of each { child} workflow. { For example, the most general \enquote{workflow of workflows} to predict the free energy gain/penalty associated with the introduction of a PTM in protein involves the following child processes:
\begin{enumerate}
    \setlength{\itemsep}{0pt}
    \setlength{\parskip}{0pt}
    \setlength{\parsep}{0pt}
    \item Protein structure inference from sequence (AlphaFold2)
    \item Parametrization of new non-standard amino acid (NWChem)
    \item Phase-space sampling through molecular dynamics (MD) simulation (GROMACS)
    \item Trajectory analysis and selection of most relevant cases (\ptmpsi\ analysis module)
    \item Thermodynamic integration for alchemical free energy calculations of selected cases (GROMACS)
\end{enumerate}
The first version of PTM-Psi was able to perform steps 1--3 in a semiautomatic fashion using the same number and node types for each step. However, the parametrization of a new non-standard amino acid in NWChem runs only on CPU and has rather large memory demands. In contrast, sampling the phase-space using MD simulations in GROMACS can be performed very efficiently in GPU-resident mode. The new version of PTM-Psi is more automatic and has better control over the resource consumption by spawning child processes that target different hardware architectures. Even better resource optimization could be achieved by letting some of the child workflows mentioned above spawn their own child workflows. For example, the preparation steps needed as prerequisites for the phase-space sampling in GROMACS can be performed in low-cost CPU-only VMs. This type of fine-grained control is an ongoing development.

To take advantage of the heterogeneous nature of cloud-based HPC solutions, we added a (Virtual) \texttt{machine} class to \ptmpsi\ where the user can specify the number of CPUs, GPUs, and memory available in each VM node. To further enhance user experience in AQE, built-in presets for several Azure VM types were also added to the \texttt{machine} class.

\ptmpsi\ was also extended with a \texttt{queue} class that can be selected to generate either SLURM or PBSPro batch files that take care of performing all the steps in a workflow and spawning child processes, while tracking the dependencies among different steps.

Finally, we included an additional module to analyze MD trajectories, especially tailored for massive datasets and large macromolecular complexes (see Subsection \nameref{sec:analysis} ).

Future enhancements to \ptmpsi\ capabilities include the addition of a tool that advises users on the best target hardware based on the specific workflow, system size, and resource availability, as well as an interface to the \texttt{Flux} management framework for better job orchestration.

All the capabilities described above are designed to minimize user intervention. However, there are still steps that have not been completely automated. For instance, the workflow to generate new forcefield parameters generates charges, and bond and angle force constants in an automatic fashion, but the user still needs to launch another process to optimize torsional potentials and, subsequently, update the necessary forcefield files manually. Once the forcefield is at hand, the rest of the thermodynamic integration workflow is completely automated, with only one command necessary to queue all necessary child workflows.

}

\subsection{Hardware Considerations and the Need for \ptmpsi\ on the Cloud}

\ptmpsi\ relies heavily on GROMACS\cite{abraham2015gromacs} and NWChem\cite{apra2020nwchem} for the most computationally demanding tasks. Both software stacks have different capabilities in terms of the hardware requirements and scalability of the algorithms being used. As a consequence, adaptation of each part of the workflows to the appropriate hardware is needed. 

{ The choice of computational platform involves trade-offs between capability, accessibility, and timeliness.
In traditional on-premise HPC clusters, homogeneity was the dominant design philosophy. This led to systems with a large number of nodes with (nearly) identical hardware. Although this model has some advantages, like simpler management and more predictable performance, it can also lead to a lack of flexibility in the software stack and/or to non-optimal performance. Moreover, deploying and expanding an on-premise HPC cluster is a complex, multi-staged project with a rather long (months to years) timeframe to complete, leading to slow procurement cycles, limited hardware diversity, and usually long queue times.} 

In contrast, HPC cloud computing is the on-demand delivery of computing power and other information technology resources via the internet with a pay-as-you-go pricing scheme. Large cloud service providers (CSP) like Microsoft Azure, Amazon Web Services, and Google Cloud Platform offer virtually limitless computing power distributed across their multi-geographic region data centers. They democratize access to a vast and varied landscape of cutting-edge hardware, bypassing the formidable entry barriers to Leadership Computing Facilities and the slowness of on-premise HPC procurement.
Both HPC, in which a large number of processors are needed to solve one particularly complex task, and high-throughput computing (HTC), in which a large number of simple and computationally independent tasks are needed, can be tackled with HPC cloud services. 
{ Since CSPs usually offer a large set of different resources,} the hardware can be configured to perfectly suit the application requirements, allowing for the perfect match between hardware and workload. { Crucially, the on-demand nature of the cloud services eliminates queueing delays--a decisive advantage for our work.}

{ The elasticity and heterogeneity of Cloud resources are ideal for the workflows implemented in \ptmpsi. For instance, while AlphaFold2 and GROMACS can be run using NVIDIA GPUs, only GROMACS officially supports AMD and Intel GPUs. We also note that non-standard residue parametrization with NWChem can only run on CPUs.
For these reasons, making \ptmpsi\ cloud-compatible, by leveraging hardware heterogeneity and exploiting the \enquote{workflow of workflows} approach, might decrease turnaround times and could potentially lead to cost savings.

As a showcase of the efficiencies that can be obtained with \ptmpsi\ on the cloud, we mention that we were able to run a full workflow of thermodynamic integration for 200 PTM combinations in the \emph{dark complex} (800k atoms) in less than 72 hours of active cloud allocation. This time includes the generation of forcefield parameters with NWChem using 20 HBv2 VMs, phase-space sampling with GROMACS using 20 NC\_H100 VMs, and the trajectory analysis and thermodynamic integration using 20 HBv2 VMs without human supervision between each child step. The NC\_H100 VMs, with a higher cost, were allocated only for 48 hours. Based on historical utilization from other on-premise clusters, we estimate that the same simulation campaign would have taken several weeks to complete.

Trajectory analysis and thermodynamic integration results for a relevant subset of the 200 combinations are presented in the following subsections.
}

\subsection{The assembly of the dark complex remains relatively stable after thiol PTMs 
}
\begin{figure}[H]
    \centering
    \includegraphics[width=1\linewidth]{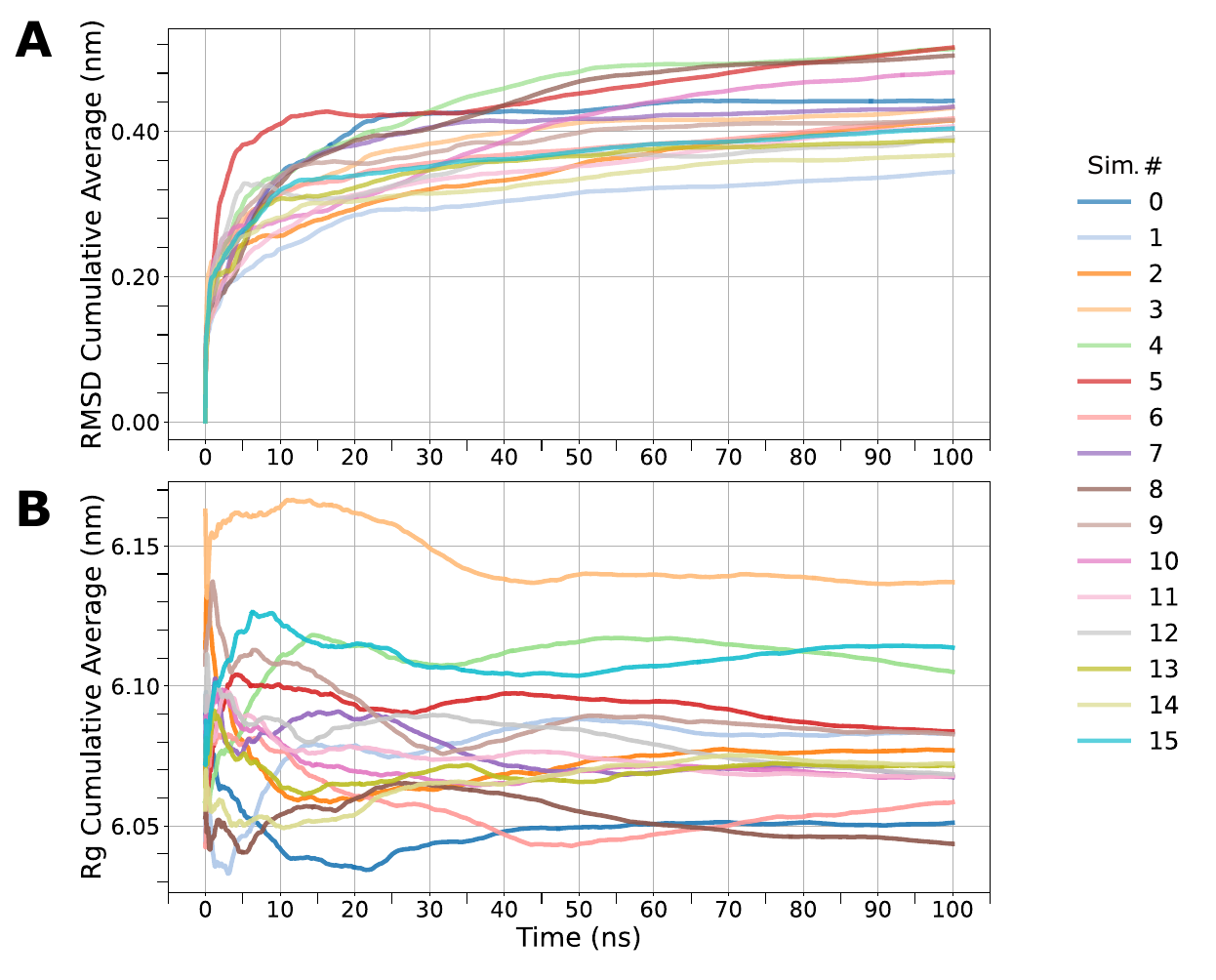}
    \caption{(A) The temporal evolution of backbone RMSD reveals a slight deviation from its starting conformation with 15 kinds of PTM perturbations described in Table \ref{table:PTM}. {Sim.}0 is the native state with two built-in disulfide bridges in PRK and two in CP12. A PTM perturbation can be either slightly oxidizing a specific cysteine of GAP2 ({Sim.}1-3), or breaking disulfide bridges by replacing them with other thiol PTMs in PRK ({Sim.}4-9) as well as CP12 ({Sim.}10-15).  The plateau reflects that conformational sampling has converged toward equilibrium during the MD simulations. (B) The \(R_g\) over time captures global dynamics reaching a steady state of consistent ternary structures with PTMed residues.}
    \label{fig:RMSD_Rg}
\end{figure}

We evaluated the global dynamics of the dark complex in its native configuration ({Sim.}0) and 15 other instances under thiol modifications as described in Table \ref{table:PTM}. We calculated the temporal evolution of RMSD as well as \(R_g\) over 100ns in Figure \ref{fig:RMSD_Rg}A and \ref{fig:RMSD_Rg}B, respectively.  {Sim.}1-3 represent the instances of the dark complex with an oxidized thiol on CYS77 with SNO, SSH, or SSG. This CYS77 is equivalent to CYS78 of the GAP2 tetramers in our previous study based on the PDB structure of \enquote{2D2I} from another cyanobacteria. CYS77 locates near the surface of GAP2 in close proximity to NAD cofactor and CP12. Its oxidation from SNO and SSH, to SSG, signifying a gradual increase in the excluded volume of the side chain. 
{Sim.}4-9 represent the reduced thiols with SNO, SSH, or SSG on the cysteine pair, breaking the constraints of disulfide bridges in PRK dimers. Likewise, {Sim.}10-15 represent the reduced thiols of a disulfide bridge in CP12, relaxing a hairpin conformation.

While the RMSD profiles vary with PTM instances, their values slowly grow while \(R_g\) profiles remain plateaued, indicating that the size of a megacomplex remains steady over time despite the structural perturbation by the inclusion of PTMs at critical sites. The molecular megacomplex maintains its ternary state, rather than undergoing significant conformational changes to a more expanded or collapsed form upon thiol PTMs within 100 ns in the trajectories.

\subsection{The Local PTM Perturbations Impact the Dynamics of Proximal Modules}

\begin{figure}[H]
    \centering
    \includegraphics[width=1\linewidth]{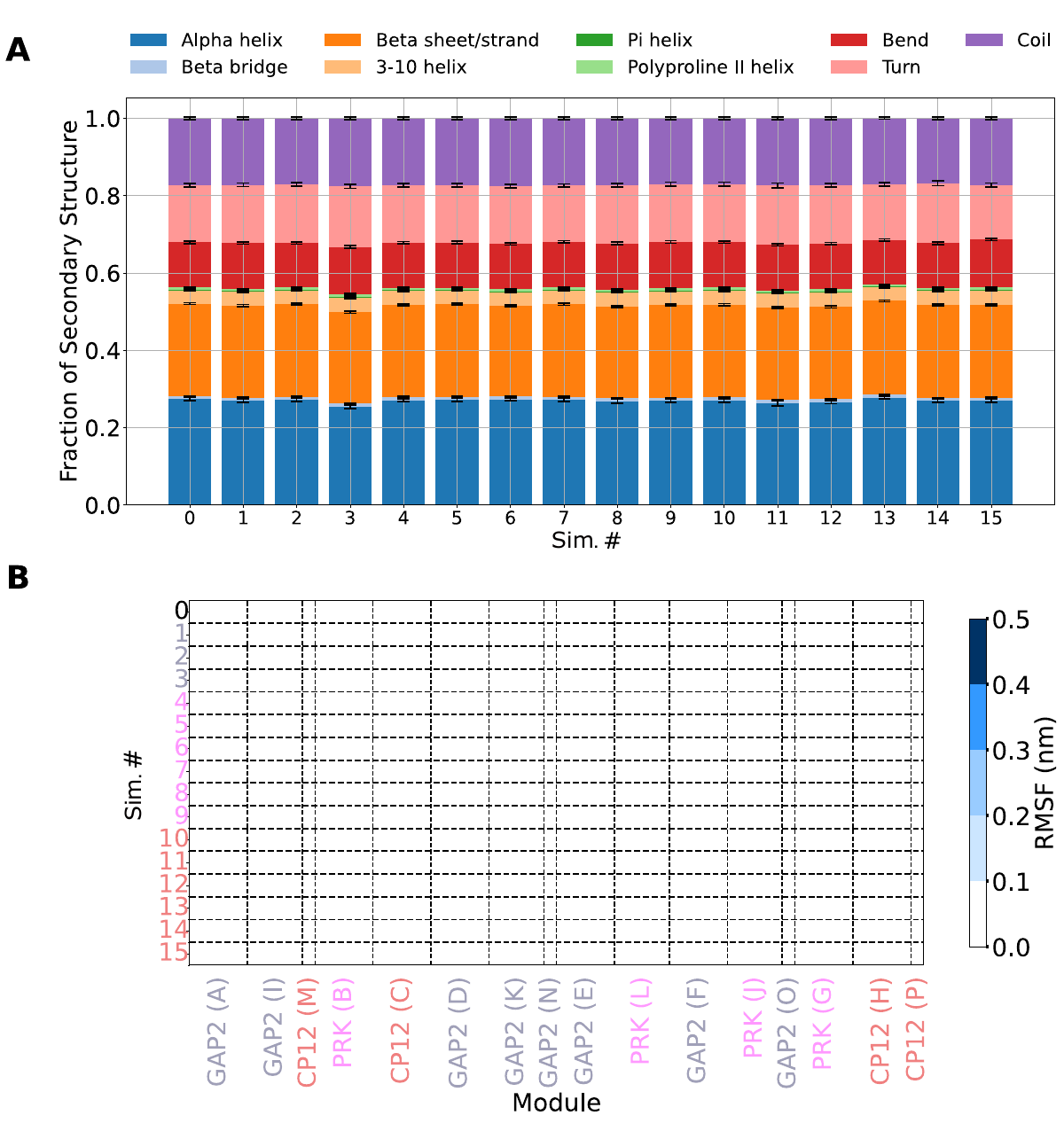}
    \caption{(A) Comparison of residue-specific secondary structure preferences averaged over trajectories under various PTM perturbations. The descriptions of the PTM perturbation for the simulation systems ({Sim.}0-15 ) can be found in Table \ref{table:PTM}. {Sim.}0 is the native configuration. (B) Comparison of per-residue RMSF analysis under various PTM perturbations. We grouped the residue indices into GAP2, PRK, and CP12 modules with a chain ID in the same color schemes as those in Figure \ref{fig:Dark_complex_schematic}. We followed the alphabetical order of the chain IDs according to their chain labels in PDB ID: 6GVE. The intensity of RMSF is provided in the color bar. RMSF greater than 0.5nm is shown in deep blue. }
    \label{fig:RMSF_SecStruct}
\end{figure}

Figure \ref{fig:RMSF_SecStruct}A show the averaged secondary structure propensity over the entire simulation under PTM perturbations. Compared to the native setting ({Sim.}0), the changes in the secondary structure motifs (e.g., $\alpha$ helix, $\beta$-sheet, coil/disordered) under various PTM perturbations are minimal, in agreement with the observation of the dark complex's global dynamics in the previous section.

When we calculated the RMSF to evaluate the residue-wise flexibility profile under PTM perturbations in \ref{fig:RMSF_SecStruct}B. 
The darker the blue color, the higher the RMSF, showing these residues are structurally flexible.  
This figure highlights the flexible or rigid regions of the dark complex under PTM perturbations. In the native configuration ({Sim.}0), Each PRK and CP12 has two pairs of disulfide bridges. Loops and termini often exhibit elevated mobility, while secondary structures such as helices and sheets remain relatively stable with low RMSF values. 

Under PTM perturbations, we observed the emergence of narrow segments with high RMSF values, not only in the regions near PTM perturbations, but also at the interface of proximal modules. For example, the RMSF of {Sim.}1, 2, and 3 (colored in gray) shows the impact of the perturbation of an oxidized cysteine from eight subunits that form two GAP2 tetrameters (Chain ID labeled with \enquote{ADEF} and \enquote{IKNO}). There are a few pronounced residual fluctuations within GAP2, as well as CP12 and PRK modules in the dark complex. When the two disulfide bridges in PRK (Chain ID labaled with \enquote{JG} and \enquote{BL}) are each perturbed with thiols, for {Sim.}4-9 (colored in magenta), a few prominent RMSF values emerge not only within PRK modules, but also near CP12 and GAP2.  Similarly, when the two disulfide bridges in CP12 are each perturbed with thiols, for {Sim.}10-15 (colored in orange), high RMSF values appear only within CP12 modules, but also near GAP2 and PRK. In summary, while we computationally perturbed PTM  one module at a time, we observed that its local impact propagates to the proximal modules of the whole dark complex.

\subsubsection{The Structural Variability of Ensemble Conformations under PTM Perturbation is revealed by PCA}

While single-variable metrics (e.g., \(R_g\) or RMSD) report isolated aspects of motion, they are unable to explain the relationships between multiple variables.  By pooling all  measurements into one
feature matrix per frame, followed by PCA, we revealed the variability of PTM perturbation projected on the first principal component. We projected  multiple features  on the first principal component, allowing us to quantitatively rank the variability caused by PTM perturbation on the conformational ensemble.

\begin{figure}[H]
\centering
\includegraphics[width=1\linewidth]{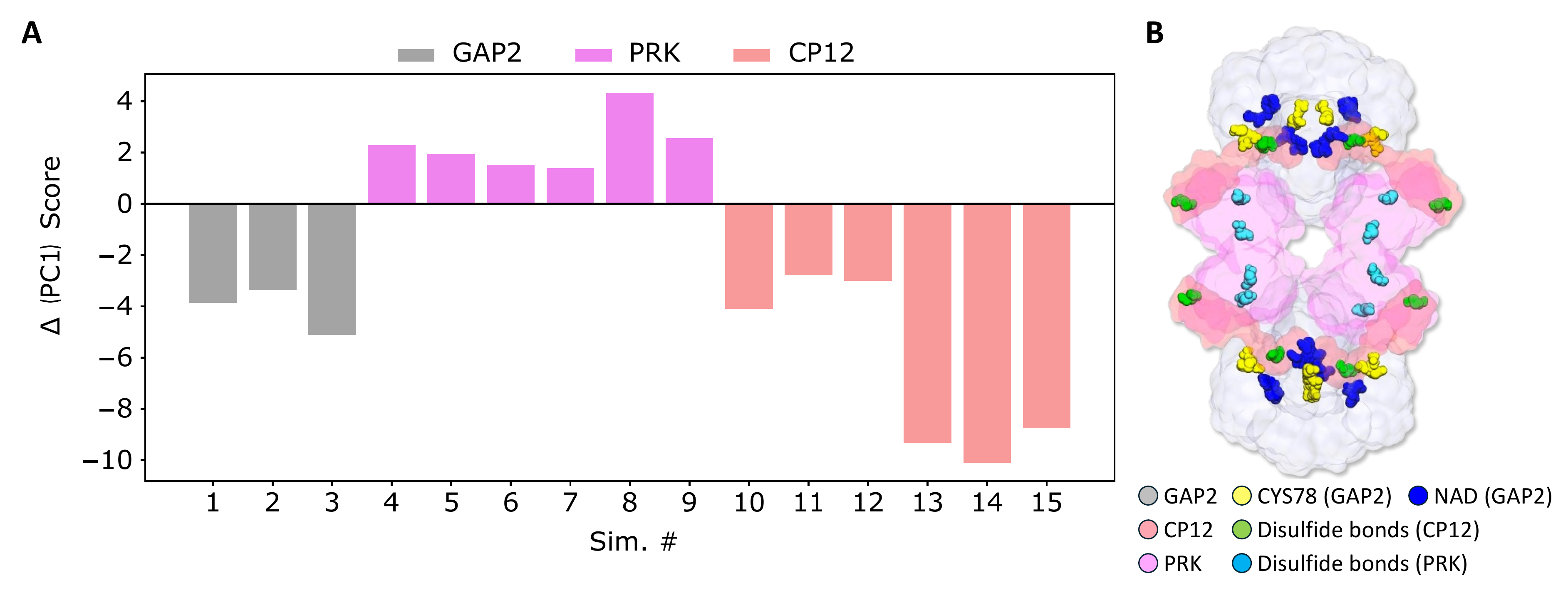}
\caption{(A) $\Delta\langle\mathrm{PC1}\rangle$ Score for each simulation system perturbed by PTMed  ({Sim.} 0 is the baseline). Positive (negative) values show positive (negative) correlation with respect to the {Sim.} 0. (B) A structural representation of GAPDH/CP12/PRK megastructure, with SSG modification in GAP2 modules of PDB: 6GVE ({Sim.} 3) in Table \ref{table:PTM}. PTM perturbations in the GAP2 modules appear in gray, PRK in pink, CP12 in magenta, oxidized cysteine (CYS78) residue of GAP2 in yellow, NAD cofactor in blue, cysteine residues forming disulfide bridges in cyan in PRK and green in CP12, respectively.}
\label{fig:pc_ranking}
\end{figure}

As detailed in \textit{Methods}, we performed a global PCA over all 16 simulations and computed the $\langle\mathrm{PC1}\rangle$. The first principal component PC1 represents the direction of maximal variance in the data. It is the axis where the data spreads out the most.  To emphasize the variability compared with the native state, using {Sim.} 0 as a baseline, we plot the $\Delta\langle\mathrm{PC1}\rangle$ score in Figure \ref{fig:pc_ranking}. It shows that the PTM perturbation drives the movement of GAP2 and CP12 opposite to the baseline, while that of PRK in the same direction as the baseline along PC1. In other words, although PTM perturbations have not altered local structures in each module significantly, they have indeed changed the global dynamics of the dark complex.

\subsection{Kinetic Transition Network}

While PCA captures the major directions of conformational variance across MD simulations, it has yet to resolve how or when a protein moves between those conformations over time. 
Therefore, we employed the KTN approach to analyze the pathways and transition states of molecular systems from the trajectories of MD by mapping the configurations onto the stable states or the transition pathways on an energy landscape.
KTN not only enables us to track the change in dynamic accessibility and the interconversion between basins on the energy landscape under PTM perturbations, but also provides the explainable features that contribute to such perturbations.

We begin by selecting the \textit{top-ranked simulation system} in Table \ref{table:PTM} from the PCA where it exhibits substantial deviation upon PTM perturbation along the PC1. 
As an example for demonstration, we selected the PTM perturbation in CYS77 of GAP2 because it is not involved in a disulfide bridge as the cysteines in other two modules. We use {Sim.}3 because it has a higher variance than {Sim.}1 and {Sim.}2. It is expected because the size of SSG PTM of {Sim.}3 is the largest among the three thiol PTMs.  We projected the feature descriptors of {Sim.}3 on the first and the second principal components, PC1 and PC2 ( accounting for $> 90\%$ of cumulative variance). We applied the $k$-means clustering of $k=3$ for three metastable states. We then constructed a Markovian transition matrix $T$ from successive frame cluster assignments and computed stationary populations $\pi$, residence times $\tau$, 
and the mean first-passage times $\mathrm{MFPT}$ for each cluster. Please see \textit{Methods} for the descriptions and definitions.

\begin{figure}[H]
\centering
\includegraphics[width=1\linewidth]{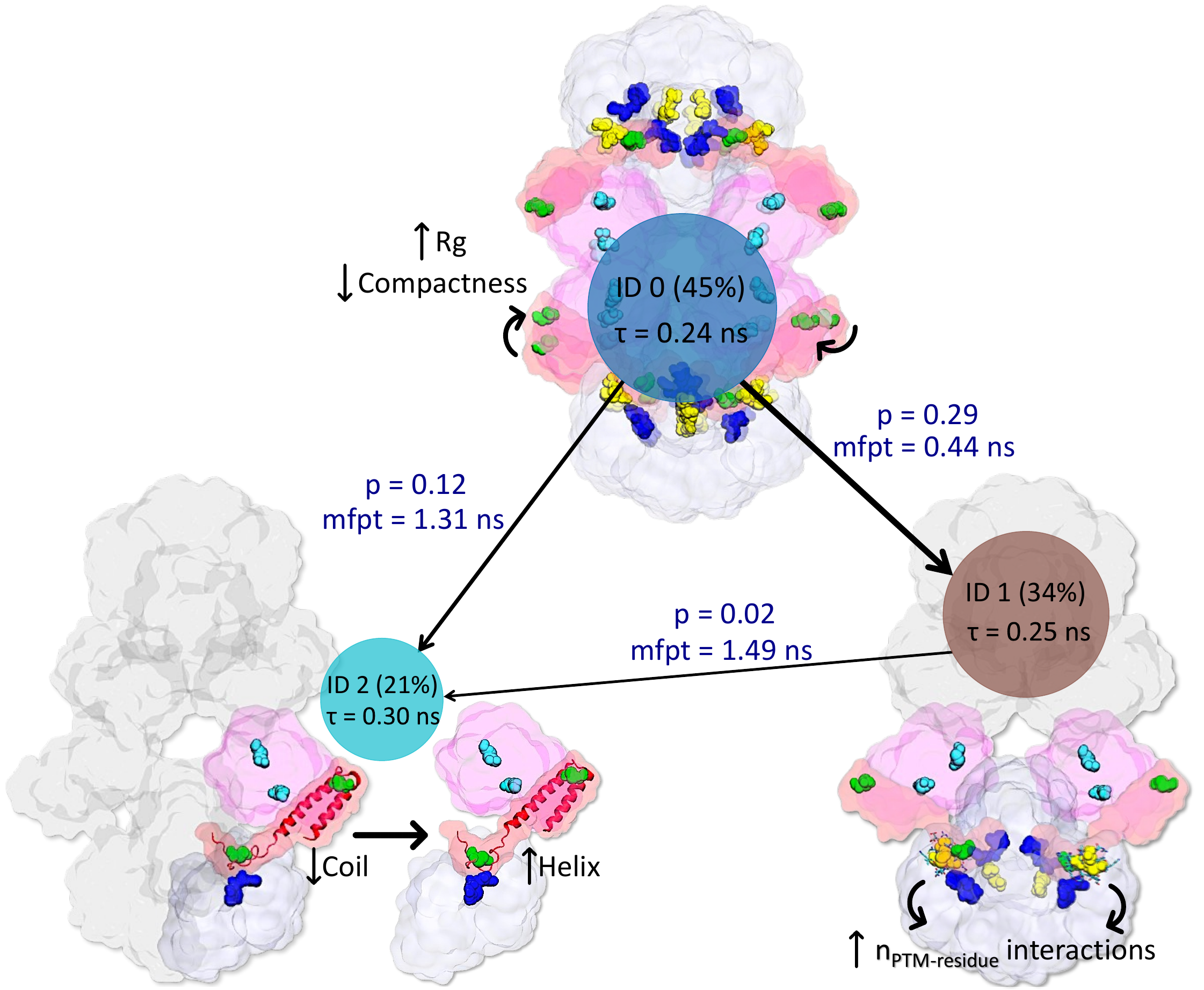}
\caption{The KTN was constructed for system {Sim.}3, where CYS77 of GAP2 was perturbed by SSG PTMs. The size of the node size is proportional to the stationary probability $\pi_j$. The width of an edge is proportional to the transition probability $T_{ij}$ (labels = $\mathrm{MFPT}_{i \to j}$ in ns). Behind each node is a representative structure near the centroid of a node. We highlighted the configurational differences in a certain region, including the loop of CP12 close to GAP2's CYS77 and NAD-binding sites. For visual guidance, NAD is colored blue, CYS77 in yellow, disulfide bridges in PRK in cyan, and CP12 in green.}
\label{fig:ktn_overlay3}
\end{figure}

Figure~\ref{fig:ktn_overlay3} shows the resultant KTN constructed from a 100\,ns trajectory of the S-glutathionylated CYS77 dark complex ({Sim.}3), revealing three metastable basins. The largest basin (cluster ID\enquote{0}) captures 45\% of the ensemble and exhibits a mean residence time, \(\tau\), of approximately 0.24\,ns. 
This cluster shows a \enquote{breathing} motion of the three GAP2, PRK, and CP12 modules, resulting in a slight increase in \(R_g\) and a decrease in compactness.
The second most populated state (cluster ID\enquote{1}) accounts for 34\% of the frames and has \(\tau\) = 0.25\,ns. 
It involves the local movement of PTMed CYS77 and nearby residues from both CP12 and GAP2 modules, increasing in $n_{PTM-residue}$.
The third basin (cluster ID\enquote{2}) comprises  21\% of the conformational ensemble and  \(\tau \approx 0.30\)~ns.  It involves helix-coil transitions in the CP12 module.

These three clusters are similarly populated, and they are each attributed to local minima separated by a kinetic barrier on an energy landscape. The probability of passing over a kinetic barrier from one minimum to another, however, varies. The most frequent transitions occur between the two basins of ID~0 and ID~1, with MFPTs of roughly 0.44\,ns. 
The PTM perturbation on CYS77 of GAP2 quickly propagates into a global \enquote{breathing} motion within the three modules in the dark complex with a relatively high transition probability of p=0.29.
In contrast, transitions from either basin ID 0 or ID 1 to basin ID 2  are relatively slow by a factor of ~4, indicating a higher barrier for  loop reorganization in CP12 than dynamics between module movements upon the perturbation of  S-glutathionylation PTM of CYS77 on GAP2.

In summary, KTN analysis not only reveals structural distinctions across the ensemble, particularly in compactness (\(R_g\)), secondary structure (coil vs.\ helix), and PTM-contact features, but also partitions the dynamics into distinct basins whose interconversion kinetics and structural signatures can be directly visualized within a single cohesive network.

%
    \label{fig:contact}


\subsection{Thermodynamic Integrations}

\begin{figure}[H]
\centering
\includegraphics[width=0.5\linewidth]{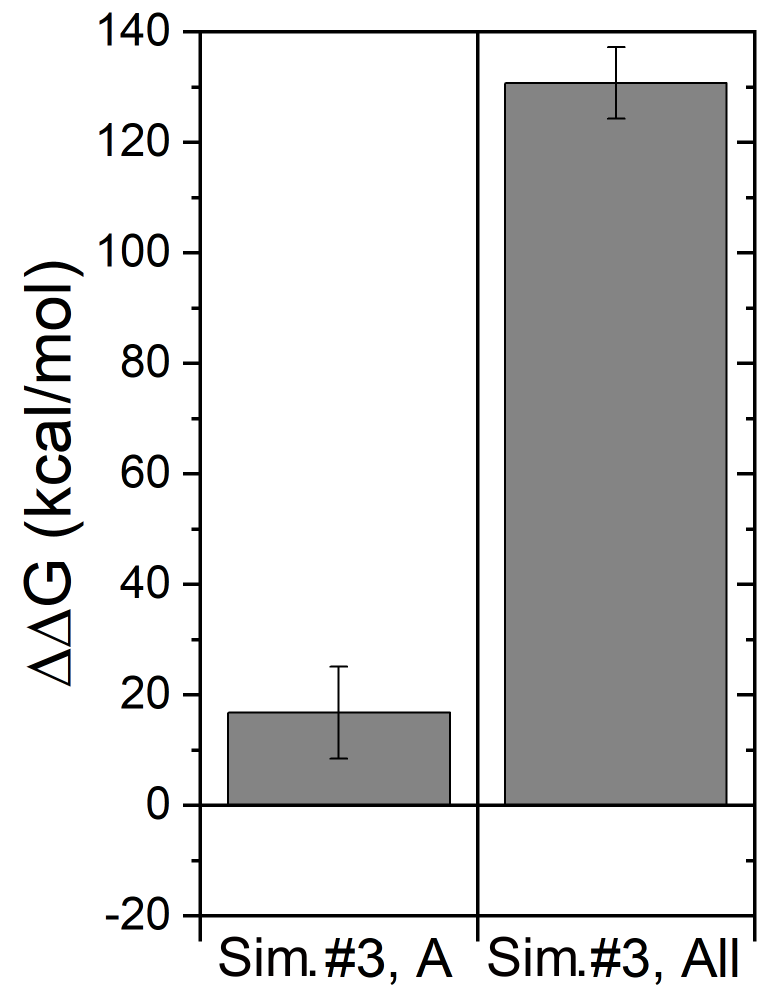}
\caption{The RBFE calculations from the TI feature in the \ptmpsi\ package. This calculation involves transforming from CYS77 in chain A of GAP2 to SSG ({Sim.}3, A) and all CYS77 residues in chain A, D, E, F, I, K, N, and O of GAP2 to SSG ({Sim.}3, All). A positive RBFE value indicates that modifications from either single or all eight CYS77 residues to SSG are energetically unfavorable in GAP2.}
\label{fig:TI_calc}
\end{figure}

One of the features of \ptmpsi\ is its ability to perform RBFE calculations using TI. We continued to focus on {Sim.}3 and calculated the RBFE to assess the energetic impact of PTMs. Specifically, we performed RBFE calculations, transforming CYS77 in chain A of GAP2 to SSG individually, as well as transforming CYS77 across all eight chains of GAP2 tetramers to SSG. As illustrated in Figure \ref{fig:TI_calc}, the RBFE calculations revealed that the PTM of a single CYS77 residue to SSG is energetically unfavorable by 16.38 kcal/mol, while the simultaneous modification of all eight CYS77 residues is even more unfavorable, with a total free energy change of 130.71 kcal/mol. It implies that SSG PTM at GAP2 might strategically destabilize the entire dark complex via propagated dynamics revealed by the KTN model. The TI analysis offers thermodynamics insights into the impact of redox on the assembly of dark complex.

\section{Conclusions}

We extended \ptmpsi\ on the cloud to make it fully scalable to a proteomic level and configurable for HPC depending on the hardware. Together with the analysis tools, including PCA and KTN that provide insight into kinetic modeling from massive data, we can further advance scientific discovery by formulating an experimentally testable hypothesis of targeting site-specific cysteines  that drive the assembly of the dark complex for redox signaling. Its far-reaching impact would be to control the yield of the CBC pathway for sugar production with the approach of synthetic biology.

\section{Conflict of Interest Statement}
The authors declare no conflicts of interest regarding this manuscript.

\section{Author Contribution}
M.S.C and D.M.-R. designed the research. P.R. and A.A. contributed to code adaptation. 
D.M.-R. and H. K. executed the simulations. S.S., H.K., and M.L. analyzed the data. All authors contributed to writing the manuscript.

\section{Data and Software Availability}

The \ptmpsi\ package is made freely available under the terms of a modified 3-clause BSD license. The source code and installation instructions can be found in the GitHub repository \href{https://github.com/pnnl/PTMPSI}{https://github.com/pnnl/PTMPSI}.
The raw data, including molecular dynamics (MD) trajectories and topology files, will be made publicly available through Pacific Northwest National Laboratory's DataHub repository.

\begin{acknowledgement}

This work is partially supported by the NW-BRaVE for Biopreparedness project funded by the U. S. Department of Energy (DOE), Office of Science, Office of Biological and Environmental Research, under FWP 81832. A portion of this research was performed on a project award (61054) from the Environmental Molecular Sciences Laboratory, a DOE Office of Science User Facility sponsored by the Biological and Environmental Research program under Contract No. DE-AC05-76RL01830. 
The research described herein was funded by the Generative AI for Science, Energy, and Security Science \& Technology Investment under the Laboratory Directed Research and Development Program at Pacific Northwest National Laboratory (PNNL). This work was also supported by the Center for AI and Center for Cloud Computing at PNNL, a multiprogram national laboratory operated by Battelle for the U.S. Department of Energy under DOE contract number DE-AC05-76RL1830.
The authors thank Song Feng and Xiaolu Li for their fruitful discussions on the redox proteomics.

\end{acknowledgement}

\bibliography{references}

\clearpage
\newpage 
\begin{tocentry}
\centering
\includegraphics[height=1.75in]{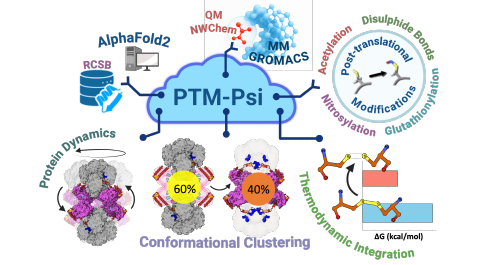}
\end{tocentry}

\end{document}